\documentclass[prl,twocolumn,showpacs,showkeys,superscriptaddress,nofootinbib]{revtex4}

\usepackage{color}
\usepackage{epsfig}

\newcommand{\beq}{\begin{equation}}
\newcommand{\eeq}{\end{equation}}
\newcommand{\ba}{\begin{array}}
\newcommand{\ea}{\end{array}}
\newcommand{\ds}{\displaystyle}
\newcommand{\beqa}{\begin{eqnarray}}
\newcommand{\eeqa}{\end{eqnarray}}
\newcommand{\beqas}{\begin{eqnarray*}}
\newcommand{\eeqas}{\end{eqnarray*}}
\newcommand{\n}{\nonumber}
\newcommand{\eps}{\epsilon}

\newcommand{\f}{\frac}

\begin{document}

\title{mKdV equation approach to zero energy states of graphene}

\author{C.-L. Ho
}
\affiliation{Department of Physics, Tamkang University, Tamsui
251, Taiwan, R.O.C.}

\author{P. Roy}
\affiliation{Physics $\&$ Applied Mathematics Unit,
Indian Statistical Institute,
Kolkata - 700 108, India}

\date{Jul 15, 2015}

\begin{abstract}

We utilize the relation between soliton solutions of the mKdV and the combined mKdV-KdV equation and  the Dirac equation to construct electrostatic fields which yield exact zero energy states of graphene.
\end{abstract}



\maketitle


{\em  Introduction.}---
The relation between solutions of nonlinear evolution equations and nonrelativistic Schr\"odinger equation is well known. For example, the soliton solutions of the KdV equation $u_t+u_{xxx}+6uu_x=0$ can be used to construct reflectionless potentials of the one dimensional Schr\"odinger equation \cite{gardiner,lamb,dodd,drazin}. There is also a similar relation between the solition solutions of the mKdV and combined mKdV-KdV equation and the Dirac equation \cite{w1,w2,w3,za,ablo,lamb}. However this relation has been exploited to a lesser extent to find out electrostatic fields admitting exact solutions of Dirac equation \cite{anderson}. 

It may be recalled that the dynamics of electrons in graphene is governed by $(2+1)$ dimensional massless Dirac equation with the exception that the velocity of light $c$ is replaced by the Fermi velocity ($v_F=c/300$) \cite{novo}. In graphene one of the important problems is to confine or control the motion of the electrons. Such confinement can be achieved, for example, by using various types of magnetic fields \cite{magnetic}. On the other hand it is generally believed that because of Klein tunneling electrostatic confinement is relatively more difficult compared to magnetic confinement. However it has been shown recently that certain types of electrostatic fields can indeed produce confinement \cite{portnoi} and zero energy states which were earlier found using various magnetic fields \cite{zero} can also be found using electrostatic fields \cite{zero1,zero2,HR}. 

In this Letter it will be shown that the Dirac equation of the graphene model is closely related to nonlinear evolution equations, namely the mKdV equation and the combined KdV-mKdV equation \cite{w3}. To be more specific, the soliton solutions of the above mentioned equations actually act as electrostatic potentials of the Dirac equation for the charge carriers in graphene. Here our objective is to use this correspondence to obtain several electrostatic field configurations which admit exact zero energy solutions of the graphene system.


{\em Zero energy states in graphene.}--- 
The motion of electrons in graphene in the presence of an electrostatic field or potential is governed by the equation
\beq\label{1}
[v_F(\sigma_xp_x+\sigma_yp_y)]\psi+U(x,y)\psi=E\psi,
\eeq
where $v_F$ is the Fermi velocity, $\sigma_{x,y}$ are the Pauli spin matrices and $U(x,y)$ is the potential. Here we would consider a potential depending only on the $x$ coordinate and consequently the wavefunction can be taken as
\beq
\psi(x)=e^{ik_yy}\left(\ba{c}\psi_A \\ \psi_B\ea\right).
\eeq
Then from Eq.(\ref{1}) we obtain
\beq\label{intert1}
\ds(V(x)-\eps)\psi_A-i\left(\f{d}{dx}+k_y\right)\psi_B=0,
\eeq
\beq\label{intert2}
\ds(V(x)-\eps)\psi_B-i\left(\f{d}{dx}-k_y\right)\psi_A=0,
\eeq
where $V(x)=U(x)/\hbar v_F$ and $\eps=E/\hbar v_F$. 

It is interesting to note that Eqs.~(\ref{intert1}) and (\ref{intert2}) are invariant under the following transformations:
\beq
k_y\to -k_y,~~ \psi_A\leftrightarrow \psi_B.
\eeq
This means that if the spinor $\psi=e^{ik_yy}(\psi_A,\psi_B)^t$ is a solution for $k_y$, then $\psi=e^{-ik_yy}(\psi_B,\psi_A)^t$ is a solution for $-k_y$ (here ``$t$" means transpose).
That is, eignestates with opposite signs of $k_y$ are spin-flipped. 

Here we shall consider $k_y\neq 0$, since for $k_y=0$ the wave function $\psi(x)$ is normalizable only in a finite region if $V(x)$ is real.
 
Defining $\psi_{1,2}=(\psi_A\pm\psi_B)$ we obtain from Eqs.(\ref{intert1}) and (\ref{intert2})
\beq\label{intert3}
\ds\left(V(x)-\eps-i\f{d}{dx}\right)\psi_1+ik_y\psi_2=0,
\eeq
\beq\label{intert4}
\ds\left(V(x)-\eps+i\f{d}{dx}\right)\psi_2-ik_y\psi_1=0.
\eeq
The above equations can be easily decoupled and the equations satisfied by the components $\psi_{1,2}$ can be obtained as
\beq\label{5}
\left[-\f{d^2}{dx^2}-(V(x)-\eps)^2-i \f{dV}{dx}+k_y^2\right]\psi_1=0,
\eeq
\beq\label{6}
\left[-\f{d^2}{dx^2}-(V(x)-\eps)^2+i \f{dV}{dx}+k_y^2\right]\psi_2=0.
\eeq

Note that Eqs.(\ref{5}) and (\ref{6}) can be interpreted as a pair of Schr\"odinger equations with energy dependent potentials which for $\eps=0$ read
\beq\label{pot3}
V_1(x)=-V^2(x)-i\f{dV}{dx}+k_y^2,
\eeq
\beq\label{pot4}
V_2(x)=-V^2(x)+i\f{dV}{dx}+k_y^2.
\eeq
It is also interesting to note that when $V(x)$ is an even function the potentials in (\ref{pot3}) or the ones appearing in Eqs.~(\ref{5}) or (\ref{6}) are $\cal{PT}$ symmetric \cite{bender}. It will now be shown that the equations for graphene's zero energy states are related to solutions of nonlinear evolution equations.


{\em mKdV equation.}---
The mKdV equation for a real field $u(x,t)$ is given by
\beq
u_t + 6u^2\,u_x+u_{xxx}=0,
\label{mKdV}
\eeq
where the subscript $t$ and $x$ indicate the time and space derivatives, respectively.
This equation is the compatibility condition of a set of linear equations \cite{lamb,w3}
\beqa
&&v_{1x} + i \zeta\,v_1=u\,v_2,\n\\
&&v_{2x} - i \zeta\,v_2=-u\,v_1.
\label{v}
\eeqa
for certain functions $v_{1,2}$.

It is seen that the transformations
\beqa
&v_1 \longleftrightarrow i\psi_B, &~~~v_2\longleftrightarrow \psi_A,\n\\
&~~\zeta \longleftrightarrow -i k_y, & u(x)\longleftrightarrow V(x).
\eeqa
map Eqs.~(\ref{intert1}) and (\ref{intert2}) into (\ref{7}) and (\ref{8}), and vice versa for $\epsilon=0$.
Accordingly, the transformation
\beqa
\phi_1 &=&-iv_1+ v_2,\n\\
\phi_2 &=&~~iv_1+ v_2,
\eeqa
changes Eq.\,(\ref{v}) into
\beqa
&&\left[-\f{d^2}{dx^2}-u^2-i u_x - \zeta^2\right]\phi_1=0,\label{7}\\
&&\left[-\f{d^2}{dx^2}-u^2+i u_x - \zeta^2\right]\phi_2=0.\label{8}
\eeqa

Comparing the two sets of Eqs.\,(\ref{7}), (\ref{8}) and (\ref{5}) and (\ref{6}), one finds that they are identical for $\epsilon=0$ with the identification $V(x)\leftrightarrow u(x), \psi_{1,2}\leftrightarrow \phi_{1,2}$ and $k_y^2\leftrightarrow -\zeta^2$.
Now the connection of  graphene's zero energy states and solutions of mKdV equation is clear.  By choosing the parameters including the time $t$ appropriately, one can make the solution of the mKdV equation an even function of $x$, i.e., $u(x)=u(-x)$.  This function $u(x)$ then furnishes a potential $V(x)=u(x)$ of the original Dirac equation (\ref{1}) admitting exactly known zero energy states. 

The soliton solutions of the mKdV equation can be obtained by the inverse scattering method. For the boundary conditions $u(x,t)\to 0$ as $|x|\to \infty$, the $N$-soliton solution is given by \cite{w2,w3}
\beq
u(x,t)=-2\frac{\partial}{\partial x}\tan^{-1} \left[\frac{{\rm Im~ det }(I+A)}{{\rm Re~ det} (I+A)}\right].
\label{u}
\eeq
where $I$ is the $N\times N$ unit matrix, and $A$ denotes the $N\times N$ matrix with elements
\beqa
A_{mn}(x,t)&=&-\frac{d_n(t)}{\zeta_n+\zeta_m}\exp\left[i(\zeta_n+\zeta_m)x\right],\n\\
                  && ~~~~m,n=1,2,\ldots, N,\n\\
     \zeta_n &=& i\eta_n, ~~~\eta_n>0,\label{A}\\
       d_n(t) &=& d_n(0) \exp(8i\zeta_n^3 t).\n
\eeqa
Here $d_n(0)$ and $\eta_n$ are real, and without loss of generality one can take
\[
\eta_1<\eta_2<\cdots<\eta_N.
\]


{\em One- and two-soliton potentials.}---
We shall present some examples of solutions of the mKdV equation that provide exactly solvable potentials for the graphene's problem. Only solutions $u(x)$ which correspond to confining potential $V(x)=u(x)$ are presented.

By taking $\zeta=i\eta,~\eta>0$, one obtains a one-soliton solution of the mKdV equation
\beq
u(x)=-2\eta\, {\rm sech}\, (2\eta x).
\label{1-soliton}
\eeq
This gives an exactly solvable potential $V(x)=u(x)$ admitting one zero energy bound state with $k_y^2=\eta^2$. The case $\eta=1/2$ correspond to the reflectionless potential with one bound state, which is a special case of the example considered in \cite{zero1,HR}.

Next, starting with two eigenvalues
\[
\zeta_1=i\eta_1,~\zeta_2=i\eta_2, ~~0<\eta_1<\eta_2.
\]
One obtains a suitable solution with two bound states for $\epsilon=0$ and $k_y^2=\eta_1^2$ and $\eta_2^2$:
\beqa\label{genpot}
&&u(x)=V(x)=4\frac{\eta_1+\eta_2}{\eta_1-\eta_2}\\
&&\times \frac{\epsilon_1\eta_1\cosh 2\eta_2x + \epsilon_2\eta_2\cosh 2\eta_1x}
{\cosh2(\eta_1+\eta_2)x + \frac{4\eta_1\eta_2\epsilon_1\epsilon_2}{(\eta_1-\eta_2)^2}+ \left(\frac{\eta_1+\eta_2}{\eta_1-\eta_2}\right)^2\cosh2(\eta_1-\eta_2)x}.\n
\eeqa
where $\eps_1,\eps_2$ are constants. This expression provides infinitely many potentials for the graphene system with two bound states. In Fig 1 we present plots of the potential (\ref{genpot}) for different values of the parameters and it is seen that depending on the parameter values the potential can be is a single well or a double well potential.

The special choice $\epsilon_1=\epsilon_2=1, \eta_1=1/2$ and $\eta_2=3/2$ gives the reflectionless potential
\beq\label{special}
u(x)=V(x)=-2\,{\rm sech}\,x.
\eeq
This corresponds to the reflectionless potential with two bound states, which is a special case of the example in \cite{zero1,HR}. 
In general, continuing the process, Eqs.~(\ref{u}) with (\ref{A}) allow one to construct potentials with $N$-bound states for the graphene system.

\begin{figure} [ht]
 \centering
 \includegraphics{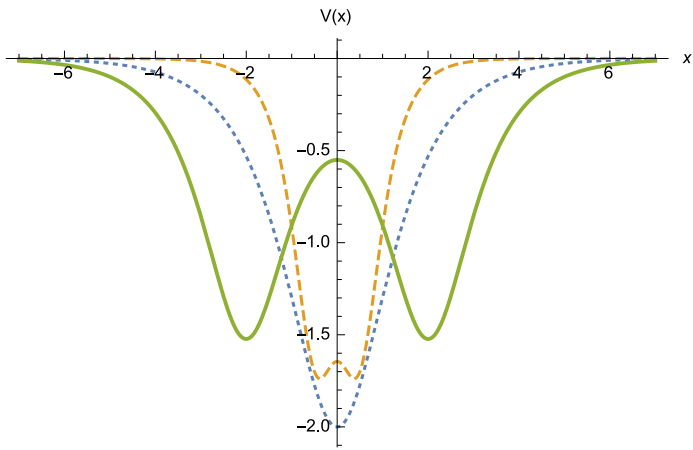}
\caption{Plot of the potential (\ref{genpot}) for  1) $\eta_1=.5,\eta_2=1.5,\eps_1=\eps_2=1$ (dotted curve); 2) $\eta_1=1,\eta_2=2,\eps_1=.5,\eps_2=.6$ (dashed curve); 3) $\eta_1=.5, \eta_2=.8,\eps_2=1.5,=\eps_2=2.6$ (thick curve).}
\label{fig1}
\end{figure}

{\em Periodic solutions.}---
The mKdV equation also admits periodic solutions \cite{w3,w4,KKSH,Fu}.
A periodic solution corresponding to one-gap state is given in \cite{w3}

\beqa\label{pot2}
&&u(x)=V(x)=\n\\
&&\frac{(a-b)(a+b+c)\,{\rm sn}^2 (\eta\, x,m)-a(a+2b+c)}{(a-b)\,{\rm sn}^2(\eta\, x,m)+(a+2b+c)},
\eeqa
where ${\rm sn}(\eta\,x,m)$ is the Jacobi elliptic function with the modulus $m (0\leq m\leq 1)$
\beq
m=\left[\frac{(a-b)(a+b+2c)}{(a-c)(a+2b+c)}\right]^\frac12,
\eeq
and the scale
\beq
\eta \equiv \frac12\left[(a-c)(a+b+c)\right]^\frac12.
\eeq
Here $a>b>c$ are constants.

This solution of mKdV equation thus furnishes a periodic potential of the graphene system that admits a zero energy bound state with $k_y^2=\eta^2$.

In the limit $m\to 1 (b=c=0)$ , ${\rm sn}\to \tanh$ and $\eta\to a/2$, and $u(x)\to -a\, {\rm sech} \,a x$, which is just the 1-solution case in Eq.~(\ref{1-soliton}). 

\begin{figure}[ht]
 \centering
 \includegraphics{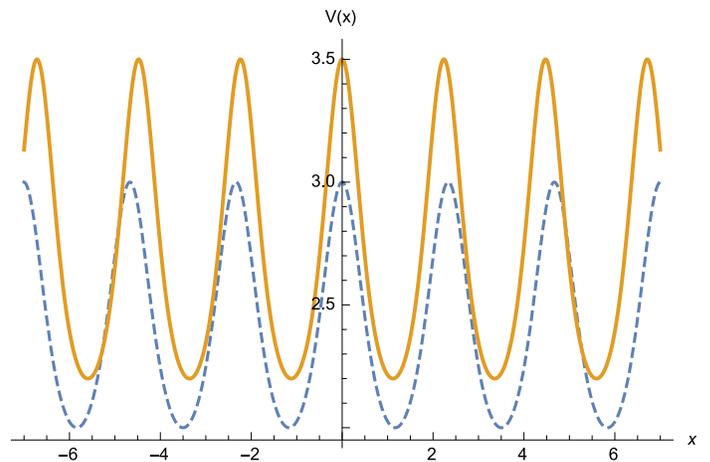}
\caption{Plot of the potential (\ref{pot2}) for 1) $a=3,b=2,c=1$ (dashed curve); 2) $a=3.5,b=2.2,c=1.4$ (thick curve).}
\label{fig2}
\end{figure}

Other periodic potentials can be obtained from the examples given in \cite{KKSH,Fu}. As an example, let us take \cite{Fu}

\beqa\label{pot3}
u(x)&=&V(x)=-m\,\eta\,{\rm cn}\,\left(\eta\,x,\, m\right);\n\\
&&~~\eta\equiv \sqrt{\frac{a}{2m^2-1}}.
\label{cn}
\eeqa

Here ${\rm cn}\, (x,m)$ is the Jacobi elliptic cosine function with modulus $m$ and $a>0$ is a constant.
For $m\to 1, {\rm cn(x,m)} \to  {\rm sech~x}$ , Eq.\,(\ref{cn}) reduces to Eq.\,(\ref{1-soliton}). In Figs 2 and 3 we present visual representations of the potentials (\ref{pot2}) and (\ref{pot3}) for different parameter values.

\begin{figure}[ht]
 \centering
 \includegraphics{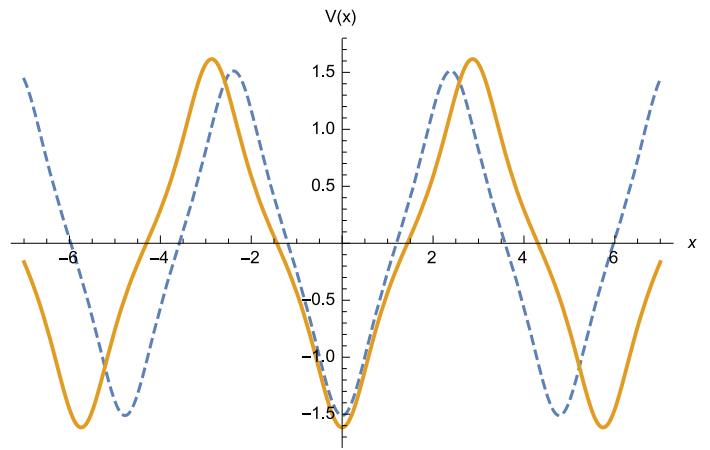}
\caption{Plot of the potential (\ref{pot3}) for 1) $m=.8,a=1$ (dashed curve); 2) $m=.9,a=2$ (thick curve).}
\label{fig3}
\end{figure}

{\em Combined KdV and mKdV equation.}---
The above observation can be extended to other soliton equations. For instance, consider the combined KdV and mKdV equation \cite{w3,w4}
\beq\label{comb}
u_t+6\alpha u u_x + 6\beta u^2 u_x + u_{xxx}=0, ~~~\beta>0.
\label{hybrid}
\eeq
It turns out that the solution $u(x)$ of this equation is connected to the potential of the Dirac equation by
\beqa
V(x)&=&\sqrt{\beta} u(x) +\gamma, ~~~\gamma\equiv \frac{\alpha}{2\sqrt{\beta}};\n\\
k_y^2&=& \eta^2 +\gamma^2.
\label{combined}
\eeqa

The $N$-soliton solutions $u(x)$ of Eq.(\ref{comb}) are known \cite{w3}.  For illustration we present the one-soliton solution
\beqa\label{pot4}
u(x)=V(x)&=&-\frac{2\eta}{\sqrt{\beta}}\,\frac{\sin\theta}{\cos\theta + \cosh 2\eta x},\n\\
\cos\theta &=& \alpha(\alpha^2+ 4\eta^2\beta)^{-1/2}.
\eeqa
Fig 4 shows that the potential (\ref{pot4}) is a single well one and the depth of the potential can be increased or decreased by choosing the parameters suitably.

\begin{figure}[ht]
 \centering
 \includegraphics{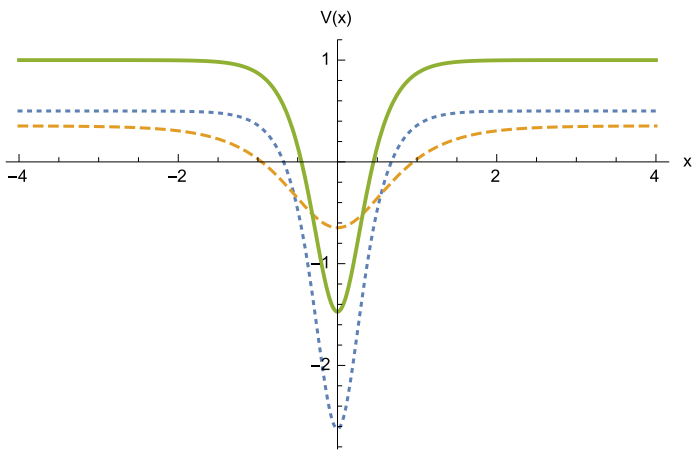}
\caption{Plot of the potential (\ref{pot4}) for 1) $\alpha=1,\beta=1,\eta=2$ (dotted curve); 2) $\alpha=1,\beta=2,\eta=1$ (dashed curve); 3) $\alpha=2,\beta=1,\eta=2$ (thick curve).}
\label{fig4}
\end{figure}


{\em Summary.}---
In this Letter we have utilized the relation between the solutions of nonlinear evolution equations, namely, the mKdV equation and combined mKdV and KdV equation and the Dirac equation to find a large number of potentials admitting exactly known zero energy solutions. There are of course many other solutions of the aforementioned equations \cite{ZZSZ} and depending on the specific need of the problem or experiment some of these solution can play an important role.  
\centerline{--------------}
\acknowledgments

The work is supported in part by the Ministry of Science and Technology (MoST)
of the Republic of China under Grant NSC-102-2112-M-032-003-MY3. 
 After the submission of our manuscript, we were informed by D.-J. Zhang of the recent review Ref.\,\cite{ZZSZ} on solutions of the mKfV equation. 



\end{document}